\documentstyle[preprint,aps]{revtex}

\newcommand\ls{\langle}
\newcommand\rs{\rangle}

\begin{document}
\title{Magnetic Charge in a Nonassociative Field Theory}
\draft
\preprint{UM-P-96/24; RCHEP-96/3}
\author{C. C. Lassig\footnote{ccl@physics.unimelb.edu.au}
and G. C. Joshi\footnote{joshi@physics.unimelb.edu.au}}
\address{Research Centre for High Energy Physics, \\
School of Physics, The University of Melbourne, \\
Parkville, Victoria 3052. Australia}
\date{April 18, 1996}
\maketitle
\begin{abstract}
The violation of the Jacobi identity by the presence of magnetic charge
is accomodated by using an explicitly nonassociative theory of octonionic
fields.
It is found that the dynamics of this theory
is simplified if the Lagrangian contains only dyonic
charges, but certain problems in the constrained quantisation remain.
The extension of these concepts to string theory may however resolve these
difficulties.
\end{abstract}

\section{Introduction}
Magnetic charge and electromagnetic duality enjoy an uncomfortable enough
status in 4-dimensional field theory, let alone from the difficulties in 
trying to extend the concept to other dimensionalities, or to string
theory \cite{Witten}. The source of much of the difficulty seems to be that
magnetic monopoles give rise to a breakdown of associativity, in the form
of an explicit 3-cocycle \cite{Jackiw}. Instead of trying to avoid this
nonassociativity, which is the usual strategy, in this paper
we attempt to include it as an essential part of the theory by using
a nonassociative algebra, that of the octonions,
to describe the potentials. 

With these nonassociative fields, it is possible to include magnetic currents
in terms of fundamental fields, rather than having to realize monopoles with
instantons, as is usually the case in field theory. It is interesting at this
point to note that such instanton solutions have been constructed in seven
and eight dimensions using octonions, satisfying a condition analogous to
a generalization of self-duality \cite{Instant}.

In Section II of this paper we examine the classical equations of motion 
that we require in order to implement our theory, and then suggest a
suitable Lagrangian density for the description of coupling to dyons, with
both electric and magnetic charge. The appropriate constrained Hamiltonian
formalism, required for quantisation, is investigated in Section III, and
various difficulties are found to arise when trying to ensure 
time-independence of the constraints. Finally, in Section IV we discuss
possible means for overcoming these problems, including using a Jordan algebra
in the context of string theory.

\section{Equations of Motion and Lagrangian Formalism}
The equations of motion of standard non-Abelian field theory can be written,
in terms of an ``electric'' current $j_{\rm e}^{\nu}$, as
\begin{eqnarray}
& & [ D_{\mu} , F^{\mu \nu} ] = j_{\rm e}^{\nu}, \label{Electric}\\
& & [ D_{\mu} , \tilde{F}^{\mu \nu} ] = 0, \label{Bianchi}
\end{eqnarray}
where $\tilde{F}^{\mu \nu}$ is the dual of the field strength tensor,
$\tilde{F}^{\mu \nu} \equiv \frac12 \epsilon^{\mu \nu \rho \sigma}
F_{\rho \sigma}$. When the field tensor and covariant derivative are 
written in terms of the potential, as $F_{\mu \nu} \equiv -[D_{\mu},D_{\nu}]
= \partial_{\mu} A_{\nu} - \partial_{\nu} A_{\mu} - [A_{\mu},A_{\nu}]$,
Eq.~(\ref{Bianchi}) is a consequence of the Jacobi identity, which would
be violated by the inclusion of magnetic sources by putting the R.H.S. equal
to some $j_{\rm m}^{\nu}$.

It is possible to accomodate such a violation if an explicitly nonassociative
algebra is used for the fields. A convenient example of such an algebra is
that of the octonions \cite{Octo}, which has a number of useful properties.
Octonions are hypercomplex numbers with seven imaginary units,
and can be expressed as a sum of these units multiplied by real components:
\begin{equation} 
x = x^0 + x^a e_a,
\end{equation}
where the units $e_a$ ($a = 1 \ldots 7$) obey the multiplication relations
\begin{equation}
e_a e_b = -\delta_{ab} + c_{abc} e_c,
\end{equation}
with the structure constants $c_{abc}$ being totally antisymmetric and equal 
to 1 for, e.g. $(abc) = (123), (145), (176), (246), (257), (347), (365)$
(different multiplication tables are possible, and a variety are used by
other authors. As far as the results of this paper are concerned, the choice is
arbitrary). The octonions, being a normed division algebra, possess a real
scalar product defined by
\begin{equation}
\ls x,y \rs = \frac12 (x \bar{y} + y \bar{x}) = x^0 y^0 + x^a y^a,
\end{equation}
which exhibits the useful properties
\begin{eqnarray}
\ls x,y \rs & = & \ls y,x \rs, \\
\ls x,[y,z] \rs & = & \ls [x,y],z \rs.
\end{eqnarray}
Unlike the other division algebras, the real and complex numbers and the
quaternions, octonions don't associate, and hence require the introduction
of an associator, analogous to the commutator:
\begin{equation}
[x,y,z] = (xy)z - x(yz).
\end{equation}
While not associative, they are however what is known as {\em alternative},
as the associator is antisymmetric under exchange of any of its arguments:
\begin{equation}
[x,y,z] = -[y,x,z] = -[x,z,y] = -[z,y,x].
\end{equation}
In such an alternative algebra the Jacobi identity becomes
\begin{equation}
[[x,y],z] + [[y,z],x] + [[z,x],y] = 6[x,y,z].
\label{Jac}
\end{equation}

If the potential $A_{\mu}(x)$ is an octonion-valued function,
Eq.~(\ref{Bianchi}) is no longer necessarily zero,
and magnetic sources are allowed:
\begin{equation}
[ D_{\mu}, \tilde{F}^{\mu \nu}] = -\epsilon^{\mu \rho \sigma \nu}
[A_{\mu}, A_{\rho}, A_{\sigma}] = j_{\rm m}^{\nu}.
\label{Magnetic}
\end{equation}

It is of course usual to obtain equations of motion from variation of an
action, and so now it falls on us to provide a suitable formalism from which
to derive Eqs.~(\ref{Electric}) and (\ref{Magnetic}). Eq.~(\ref{Electric})
on its own can naturally be obtained from the Lagrangian
\begin{equation}
{\cal L}_{\rm e}(x) = -\frac14 \ls F_{\mu \nu},F^{\mu \nu} \rs
- \ls A_{\mu}, j_{\rm e}^{\mu} \rs.
\label{Le}
\end{equation}
The scalar product here ensures that the action is real, and performs the
same function as the trace over generators in non-Abelian field theory.

As for the dual equation, Eq.~(\ref{Magnetic}), it can be derived by using
a Lagrangian density
\begin{equation}
{\cal L}_{\rm m}(x) = -\frac14 \ls F_{\mu \nu}, \tilde{F}^{\mu \nu} \rs
- \ls A_{\mu}, j_{\rm m}^{\mu} \rs.
\label{Lm}
\end{equation}
When expanded in the potentials we find that
\begin{equation}
\ls F_{\mu \nu}, \tilde{F}^{\mu \nu} \rs
= - \epsilon^{\mu \nu \rho \sigma} 
\ls A_{\mu}, [A_{\nu},A_{\rho},A_{\sigma}] \rs
+ \mbox{total divergence}.
\label{Div}
\end{equation}
In the standard non-Abelian theory, with associative fields, this entire
term can be left out of the Lagrangian as a total divergence doesn't 
contribute to the action integral. With nonassociative fields however,
it persists, and allows a coupling to magnetic sources.

Now we have two Lagrangians, an electric and magnetic, but we would like only
one to satisfy our dynamical requirements. Hence we incorporate both 
couplings into a single action, meaning that our sources
are dyonic, with electric and magnetic charge.
The total Lagrangian is then
\begin{equation}
{\cal L}_{\rm em}(x) = -\frac14 \ls F_{\mu \nu},F^{\mu \nu} \rs
-\frac14 \ls F_{\mu \nu}, \tilde{F}^{\mu \nu} \rs - \ls A_{\mu}, j^{\mu} \rs.
\label{Lem}
\end{equation}
The consequence of using this ``electromagnetic'' formalism is that now we
only have one equation of motion for both kinds of coupling:
\begin{equation}
[D_{\mu}, F^{\mu \nu}] + [D_{\mu}, \tilde{F}^{\mu \nu}] = j^{\nu}.
\label{EM}
\end{equation}
In constructing Eq.~(\ref{Lem}) we have arbitrarily chosen equal strength
couplings for the electric and magnetic terms, but this is not necessarily
so, and a relative strength to the two terms should be included. At this
point however we do not know what that relative strength is, as our theory
has not provided us with an equivalent to the Dirac quantisation condition.
In the next section we will see that the choice of equal couplings in fact
simplifies the dynamics somewhat.

\section{Hamiltonian Formalism and Quantisation}
In this section we will follow the standard procedure for quantisation of
constrained systems \cite{Dirac}.
For simplicity, we will at first consider only the purely ``electric''
dynamics of Eq.~(\ref{Le}), but without source terms.
The effect of the magnetic terms will be included later.
Using our potentials $A_{\mu}(x)$ as the configuration
variables, the conjugate momenta are the electric fields, 
$E^{\mu}(x) \equiv F^{\mu 0}(x)$. With the primary constraint 
$E^0 (x) = 0$ included with a Lagrange multiplier, the total Hamiltonian is
\begin{eqnarray}
H_T & = & H + \int d^3x \ls \lambda, E^0 \rs \nonumber \\
& = & \int d^3x \Big( \frac12 \ls E^i,E^i \rs + \frac14 \ls F_{ij} , F_{ij} \rs
- \ls A_0 , \partial_i E^i + [E^i,A_i] \rs + \ls \lambda, E^0 \rs \Big) .
\label{HT}
\end{eqnarray}
Using Hamilton's Principle, from the variation of the Hamiltonian with respect
to the conjugate position and momenta, we obtain the equations of motion:
\begin{eqnarray}
\frac{dA_i}{dt} & = & E^i + \partial_i A_0 - [A_i,A_0], \\
\frac{dE^i}{dt} & = & -\partial_j F_{ij} + [A_j,F_{ij}] - [E^i,A_0], \\
\frac{dA_0}{dt} & = & \lambda, \\
\frac{dE^0}{dt} & = & \partial_i E^i + [E^i,A_i].
\end{eqnarray}
Because of the nonassociativity of the fields, we are not using a Poisson
bracket notation. If however we expand all octonions into their real
components, fully associative fields are obtained which can be treated as
standard position and momenta. 
In the process of quantisation, then, if the operators are chosen
to correspond to the real components of the fields, then 
they will associate, and be compatible with standard quantum mechanics
--- a nonassociative generalisation is not required. The nonassociativity
of the original fields is then reduced to a property of the structure
constants which will appear in the expansion of $F_{\mu \nu}$.
Any results obtained using such a notation will be equivalent to those
derived here.

To keep the primary constraint satisfied at all times, an
appropriate secondary constraint
$\Gamma~\equiv~\partial_i E^i~+~[E^i,A_i]~=~0$ is introduced.
This constraint must also be time-independent, but we find its variation to be
\begin{equation}
\frac{d\Gamma}{dt} = 3 [F_{ij},A_i,A_j] + [A_0,\Gamma]
+ 6 [A_0,E^i,A_i],
\label{dGdt}
\end{equation}
in the derivation of which we have used the alternativity property
Eq.~(\ref{Jac}). Up to this point, the dynamics have been the same as for
non-Abelian field theory \cite{ItZ},
except that the rate of change in $\Gamma$ usually
appears without the associator terms of Eq.~(\ref{dGdt}), in which case
$\Gamma$ is time-independent under the constraint conditions. However for our
non-associative theory we require yet another constraint,
\begin{equation}
\Lambda \equiv 3 [F_{ij},A_i,A_j] + 6 [A_0,E^i,A_i] = 0.
\label{Lambda}
\end{equation}
When we check the time-dependence of this constraint, we find
\begin{eqnarray}
\frac16 \frac{d\Lambda}{dt} & = &
[\partial_i E^j,A_i,A_j] - [[\partial_i A_j,A_0],A_i,A_j]
- [[E^i,A_j],A_i,A_j] + [[[A_i,A_0],A_j],A_i,A_j] \nonumber \\
& & - [E^i,F_{ij},A_j] + [F_{ij},\partial_i A_0,A_j] - [F_{ij},[A_i,A_0],A_j]
+ [A_0,E^i,\partial_i A_0] \nonumber \\
& & - [A_0,E^i,[A_i,A_0]] - [A_0,\partial_j F_{ij},A_i]
+ [A_0,[A_j,F_{ij}],A_i] - [A_0,[E^i,A_0],A_i] \nonumber\\
& & + [\lambda,E^i,A_i].
\label{dLdt}
\end{eqnarray}
It is possible to simplify this equation slightly through various algebraic
manipulations, but not to any great advantage. It is also possible to 
include the sourceless form of Eq.~(\ref{Magnetic}), which is fair as we
have used the electric equivalent, that requires that the potentials
associate, i.e. form a quaternionic subalgebra. It does not, however,
guarantee that the derivatives associate, and so the associators of 
Eq.~(\ref{dLdt}) don't all vanish under this condition. The most we can
say is that the sourceless theory is {\em locally} quaternionic.
(We could enforce the restriction to the quaternion subalgebra globally,
i.e. have essentially an SU(2) field theory,
and introduce nonassociativity at the quantum level,
but this isn't general enough for our requirements).
Thus we cannot reduce Eq.~(\ref{dLdt}), and there doesn't seem to be any
$\lambda$ that will give us $d\Lambda/dt~=~0$, and so complete this
Hamiltonian formalism and allow a satisfactory quantisation of the theory.

Now we look at what happens if we start with an electromagnetic Lagrangian,
with the magnetic term included with a coupling constant $\kappa$ relative
to the electric term. If we expand in the manner of Eq.~(\ref{Div}), and
integrate over the divergence, it only contributes a velocity-independent
interaction term to the Hamiltonian. We can however retain its full form,
and find that the conjugate momenta are now
\begin{equation}
P^{\mu}(x) \equiv F^{\mu 0}(x) + \kappa \tilde{F}^{\mu 0},
\label{NewP}
\end{equation}
i.e. the sum of the electric and magnetic fields. The new Hamiltonian
then becomes
\begin{equation}
H^{\prime} = \int d^3x \Big( \frac12 \ls P^i,P^i \rs 
+ \frac14 (1 - \kappa^2) \ls F_{ij} , F_{ij} \rs
+ \frac12 \kappa \epsilon^{ijk} \ls P^i , F_{jk} \rs
- \ls A_0 , \partial_i P^i + [P^i,A_i] \rs \Big) .
\end{equation}
Inspection of this Hamiltonian reveals that the dynamics can be simplified
by the choice of the coupling parameter, so henceforth we will put $\kappa = 1$,
as we did initially in our {\em ad hoc} construction. 

The constraint structure is similar to the electric case, and results in the
final constraint
$\Lambda^{\prime} = 3 \epsilon^{ijk} [P^i,A_j,A_k] + 6 [A_0,P^i,A_i]$, with
\begin{eqnarray}
\frac16 \frac{d\Lambda^{\prime}}{dt} & = &
\epsilon^{ijk} \Big( \frac12 [[A_0,P^i],A_j,A_k] + [P^i,\partial_j A_0, A_k]
+ [P^i,[A_0,A_j],A_k] \nonumber \\
& & + [A_0,\partial_j P^i,A_k] + [A_0,[P^i,A_j],A_k] + \frac12 [A_0,P^i,F_{jk}]
\Big) \nonumber \\
& & + [\partial_j P^i,A_i,A_j] + [[P^i,A_j],A_i,A_j] - [P^i,F_{ij},A_j] 
\nonumber \\
& & + [A_0,P^i,\partial_i A_0] + [A_0,[A_0,P^i],A_i] + [A_0,P^i,[A_0,A_i]]
\nonumber \\
& & + [\lambda,P^i,A_i].
\label{dL'dt}
\end{eqnarray}
Again it doesn't appear to be possible to cancel this equation by a choice
of $\lambda$.

\section{Conclusions and Further Possibilities}
In this paper we have found a means to include magnetic charge in a field
theory by using a nonassociative algebra. The most natural way of doing this
seems to be by using an ``electromagnetic'' Lagrangian, Eq.~(\ref{Lem}),
in which the charges are in fact dyonic, with equal electric and magnetic
couplings to the gauge fields. Unfortunately it is not possible to construct
a consistent constrained Hamiltonian dynamics from this Lagrangian, which
becomes an obstacle to the quantisation of the theory. Another, more aesthetic
problem is that we would like to have separate electric and magnetic charges,
but this would seem to entail having two separate Lagrangians, Eq.~(\ref{Le})
and Eq.~(\ref{Lm}). It is interesting to note that one solution may solve
both these problems.

Firstly, it seems to be the very nonassociativity which we desired that
prevents the construction of a successful field theory, and yet nonassociative
structures have been productively applied in the use of
the $M_3^8$ Jordan algebra, of $3 \times 3$ matrices of octonions, in
heterotic string theory \cite{String}.
The ideas developed in this paper then, while not
seeming to work in 4-dimensional octonionic field theory,
may instead find a home in 10-dimensional superstring theory.

As for the problem with incorporating two equations of motion, 
into a Lagrangian or
Hamiltonian formalism, Nambu has conceived of a ``generalised Hamiltonian
dynamics'', in which a canonical triplet of three dynamical variables has
equations of motion derived from two Hamiltonians, using a generalisation
of the Poisson bracket with three arguments. At the classical level this
method can be used to describe the motion of a rotator, but difficulties
arise in quantisation when trying to find the operator equivalent of
the Nambu bracket. Most of the possibilities
considered by Nambu either fail to match his requirements, or are
equivalent to normal one-Hamiltonian dynamics. The Jordan algebra $M_3^8$,
however, has potential as a candidate to satisfy the algebraic requirements. 

Thus a possibility worth considering is the use of the $M_3^8$ Jordan algebra,
most probably as applied to the 10-dimensional heterotic string, 
which may provide an implementation
of the dual equations of motion through dual Hamiltonians, and likely
a working quantum field theory with fundamental magnetic charges.

\end{document}